\documentclass[twoside,prl,twocolumn,amsfonts,amssymb,amsmath,lengthcheck,showpacs]{revtex4}
\usepackage{times}
\usepackage{latexsym}
\usepackage{bm}
\usepackage{textcomp}
\usepackage{graphicx}
\usepackage{amsmath}
\usepackage{amsfonts}

\newcommand{\beq}{\begin{equation}}
\newcommand{\eeq}{\end{equation}}
\newcommand{\bec}{\begin{center}}
\newcommand{\eec}{\end{center}}
\newcommand{\braket}[1]{\ensuremath{\left\langle#1\right\rangle}}
\renewcommand{\mod}[1]{\ensuremath{\left|#1\right|}}

\renewcommand{\subsection}[1]{\vspace{5pt}{\centering\bf #1\\}\vspace{5pt}}

\begin{document}

\title{Stability of Bosonic atomic and molecular condensates near a Feshbach resonance}
\date{July 19, 2005}
\author{Sourish Basu}
\email{sourish@ccmr.cornell.edu}
\author{Erich J. Mueller}
\email{emueller@ccmr.cornell.edu}
\affiliation{Laboratory of Atomic and Solid State Physics, Cornell University, Ithaca, New York 14853}
\begin{abstract}
We explore the Bose condensed phases of an atomic gas on the molecular side of a Feshbach resonance.  In the absence of atom-molecule and molecule-molecule scattering, we show that the atomic condensate is either a saddle point of the free energy with energy always exceeding that of the molecular condensate, or has a negative compressibility (hence unstable to density fluctuations). Therefore no phase transition occurs between the atomic and molecular condensates. We also show that a repulsive molecule-molecule scattering can stabilize a sufficiently dense atomic condensate, leading to the possibility of a phase transition. We caution that 3-body processes may render this transition unobservable.
\end{abstract}
\maketitle
Using Feshbach resonances \cite{feshbach:feshbach}, experimentalists can tune the interactions in atomic clouds\cite{inouye:feshbach, donley:coherence, strecker:fermi, ohara:fermi, greiner:fermi, chin:pairing, bourdel:becbcs}. For a system of Fermi atoms, this technique has allowed the study of a crossover between a BCS superfluid of Cooper pairs to a BEC superfluid of molecules \cite{ohara:fermi, greiner:fermi, chin:pairing, bourdel:becbcs}. Recently three separate theoretical groups\cite{romans:qpt, radzihovsky:qpt, lee:qpt} have proposed that for Bosonic atoms the same technique can produce a phase transition between an atomic and a molecular superfluid (respectively called ASF and MSF henceforth, after \cite{radzihovsky:qpt}).  If, as suggested by Romans {\it et al} \cite{romans:qpt} and Radzihovsky {\it et al} \cite{radzihovsky:qpt}, this quantum transition is continuous, then it would be in the Ising universality class, with dramatic signatures in the properties of vortices.  The topological character of this phase transition makes it of intense interest to a large community of physicists.

Here we show that in the limit of vanishing molecule-molecule and atom-molecule interaction (the same limit considered in \cite{romans:qpt} for constructing their phase diagram), no ASF\(\leftrightarrow\)MSF phase transition can occur near a Feshbach
resonance (defined as where the molecular binding energy approaches zero). Previous work\cite{romans:qpt, radzihovsky:qpt} showed that as one decreases the magnitude of the molecular binding energy, but before reaching resonance, the MSF becomes unstable.  In those works it was {\em assumed} that the instability leads to a phase containing atomic superfluid order. We demonstrate that without the stabilizing influence of molecule-molecule interaction the system has a negative compressibility and this instability actually leads to a mechanical collapse of the cloud, and adding repulsive molecule-molecule interaction only stabilizes the cloud at sufficiently high density (estimated at least three orders of magnitude higher than current experiments \cite{claussen:dynamics, donley:coherence, xu:msf}).

Experiments are routinely performed \cite{donley:coherence, claussen:dynamics} on dilute atomic clouds on the molecular side of a
resonance (i.e., the side on which a bound molecule exists), so this instability cannot be the whole story. Indeed, we verify the existence of a mechanically stable ASF in this region and show that it always has a larger energy than the molecular condensate, precluding the possibility of a phase transition (even a first order one). Furthermore, we demonstrate that this ASF is a saddle point of
the free energy, and it is always energetically favorable for atoms to recombine into molecules. It is only the slow kinetics of this
recombination, which relies upon three-body collisions, which allows experiments to be performed on atomic condensates.

\subsection{Phase diagram}
We model the Hamiltonian for a mixture of atoms and
molecules near a one-channel Feshbach resonance as
\begin{align}
&\mathcal{F} = \int
\left[F_m(x)+F_a(x)+F_{am}(x)\right]d^3x\label{eq:freeenergy}\\
&F_{a} = \frac{\nabla \psi^\dagger_a \nabla \psi_a}{2m}-\mu
\psi^\dagger_a \psi_a + \frac{\lambda_a}{2}\psi^\dagger_a\psi^\dagger_a\psi_a\psi_a\notag\\
&F_{am} = g \left[\psi_m^\dagger
  \psi_a\psi_a+\psi_a^\dagger\psi_a^\dagger\psi_m\right]+\lambda_{am}
  \psi^\dagger_m\psi^\dagger_a\psi_a\psi_m\notag\\ 
&F_{m} = \frac{\nabla \psi^\dagger_m \nabla
\psi_m}{4m}+(\epsilon-2\mu)\psi^\dagger_m
  \psi_m+\frac{\lambda_m}{2}\psi^\dagger_m\psi^\dagger_m\psi_m\psi_m\notag
\end{align}
where $F_a$ and $F_m$ represent the pure atomic and molecular contributions, and $F_{am}$ the coupling between them.  Field operators $\psi_a(x)$ and $\psi_m(x)$ respectively annihilate atoms and molecules at position $x$ (which is suppressed in these
equations). Parameters $\lambda$ represent the strengths of elastic scattering, while $g$ represents the strength of conversion between atoms and molecules, $\mu$ is the chemical potential, and \(\epsilon<0\) is the binding energy of a molecule, which can be
controlled by tuning an external magnetic field.  To treat this Hamiltonian within mean field theory one must renormalize the coupling
constants from their bare values. For example, Duine and Stoof\cite{duine:renormalization} have derived a simple renormalization scheme which connects these quantities with their bare values, providing their magnetic field dependence.
\begin{figure}[tbhp]
\includegraphics[width=\columnwidth,clip]{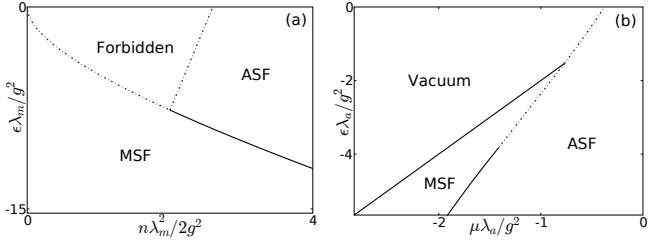}
\caption{\label{fig:phasediagram}Phase diagrams of Eq. (\ref{eq:freeenergy}) without background atom-molecule scattering (\(\lambda_{am}=0\)) in the parameter space of binding energy \(\epsilon\) and density \(n\) (a) or chemical potential \(\mu\) (b), at \(\lambda_m=2\lambda_a\). The various dotted lines separate phases with a discontinuous transition, while the solid lines denote a continuous transition. The tricritical point between MSF and ASF in the two figures, at \(n\lambda_a\lambda_m=2g^2\), corresponds to the lowest density for which a stable ASF exists.}
\end{figure}

In this letter, we find the stationary points of \eqref{eq:freeenergy}, and analyze their stability. We discuss two types of stability: dynamic, where small fluctuations do not grow in time; and thermodynamic, where small fluctuations
cannot reduce the free energy. Although a thermodynamic instability implies that the system will eventually decay, the timescale, which is governed by kinetics and dissipation, may be long enough that the system appears stable.  (In fact, since the ground state of alkali atoms at nano-Kelvin temperatures is a solid, all experiments on ultracold atoms involve states which are thermodynamically unstable.) Following convention, we take a thermodynamically unstable (but dynamically stable) phase to be metastable.

As shown by previous authors\cite{romans:qpt,radzihovsky:qpt}, for \(g\neq0\), there are two possible superfluid orders: (a) a pure molecular condensate $\phi_m=\braket{\psi_m}\neq0$, $\phi_a=\braket{\phi_a}=0$; and (b) a mixed atomic/molecular condensate $\phi_a\neq0$, $\phi_m\neq0$. States with these respective orders will be called a molecular superfluid (MSF) and an atomic superfluid (ASF).

Generically there are two classes of modes which can destabilize these states: density fluctuations, and pairing fluctuations. The latter
modes change the relative population of atomic and molecular states without changing the total density. Mueller and Baym \cite{mueller:collapse} characterized both types of modes within a random phase approximation, showing that in the absence of a molecular bound state there is no phase transition between an atomic and paired superfluid.  Our current calculation extends this result to the case where a true molecular bound state exists. 

Our primary result is the phase diagram in figure~\ref{fig:phasediagram}, shown for \(\lambda_{am}=0\) and \(\lambda_m/\lambda_a=2\). Due to the presence of metastability in these experiments we do not limit our discussion to the thermodynamic ground state in each region, but also analyze the stability of other stationary points of the energy, which can have either ASF or MSF character. These stationary points, hereafter called solutions, can be found by working at either fixed density or fixed chemical potential.

Fixing the density, the ``forbidden'' region\footnote{Such forbidden regions, corresponding to coexistence of two bulk phases, are generic features of first order phase transitions.} contains three or four solutions: A\(_1\), an ASF thermodynamically unstable to pairing; A\(_2\), an ASF dynamically unstable to density fluctuations; M, an MSF unstable to pairing; and optionally A\(_3\), an ASF dynamically unstable to relative phase fluctuations. The ``MSF'' region contains a stable MSF, and either one (A\(_1\)) or three (A\(_0\), A\(_1\), A\(_2\)) ASF solutions, two of which (A\(_0\), A\(_1\)) are thermodynamically unstable to pairing while the other (A\(_2\)) is dynamically unstable to density fluctuations. A\(_1\), however, is \emph{dynamically} stable against all fluctuations if \(\epsilon\lambda_a<2g^2\). That is, under these conditions, A\(_1\) is metastable. The ``ASF'' region contains two (A\(_1\), A\(_2\)) or three (A\(_1\), A\(_2\), A\(_3\)) ASF solutions, one (A\(_2\)) of which is stable. In this region the MSF (M) is unstable to pairing.

Fixing the chemical potential, the ``vacuum'', where the ground state contains no particles, has an unphysical ASF solution\footnote{It corresponds to negative atomic density.}. The ``MSF'' region contains a stable MSF solution, an unphysical ASF solution\cite{endnote21}, and possibly two more ASF solutions, one unstable and the other is metastable, possessing a higher free energy than the MSF. The ``ASF'' region contains three ASF solutions, one of which is unphysical\cite{endnote21}, one unstable\footnote{This solution is stabilized for \(\mu>0\) (not shown).}, and one stable. The MSF is either unstable to pairing or has a higher free energy than this ASF solution.

In the remainder of this paper, we derive these results; we find the stationary states of the Hamiltonian~\eqref{eq:freeenergy} and analyze their dynamic and thermodynamic stability against the two forms of fluctuation at \(\lambda_m=\lambda_{am}=0\). We then explore the role of finite \(\lambda_m\) and \(\lambda_{am}\). We give full details for the calculation at fixed density, and briefly sketch the procedure for fixed chemical potential.

\subsection{Stationary States (fixed density)}
\begin{figure}[tbhp]
\includegraphics[width=\columnwidth,clip]{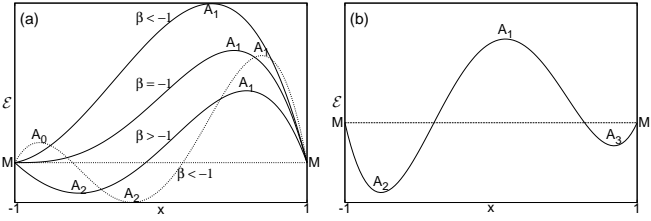}
\caption{\label{fig:energy} Scaled energy $\mathcal{E}=\braket{\mathcal{F}}+(\mu-\epsilon/2)n$ versus molecular condensate order parameter $x=\phi_m \sqrt{2/n}$ for fixed $n$. (a) \(\lambda_m=\lambda_{am}=0\): curves show $-\epsilon>2g\sqrt{2 n}$ (\(\beta<-1\)), $-\epsilon<2g\sqrt{2 n}$ (\(\beta>-1\)), and $-\epsilon=2g\sqrt{2 n}$ (\(\beta=-1\)). For \(\beta<-1\) there are two sub-cases: either a single stationary point (solid line) or three stationary points (dotted line). (b) \(\lambda_m,\lambda_{am}\neq0\): A\(_3\) does not exist if both of them are zero.}
\end{figure}

Assuming a uniform condensate exists, we replace the field operators in Eq.~(\ref{eq:freeenergy}) by their expectation values, \(\phi_m=\braket{\psi_m}=\sqrt{n_m}e^{i\theta_m}\) and \(\phi_a=\braket{\psi_a}=\sqrt{n_a}e^{i\theta_a}\), where $n_{a/m}$ and $\theta_{a/m}$ are the number of condensed atoms/molecules and their phase. The energy only depends upon the phase difference $4\xi=\theta_m-2\theta_a$, so without any loss of generality we will take $\phi_a$ to be real and positive.  Setting  $\partial \langle\mathcal{F}\rangle/\partial\xi=0$ shows that $\phi_m$ must also be real, but not necessarily positive. We work at fixed density, $n=n_a+2n_m$, writing $\phi_m=\sqrt{n/2} x$, and $\phi_a=\sqrt{n}\sqrt{1-x^2}$ with $-1\leq x\leq 1$.  The points $x=\pm1$ represent the same state.  The shifted energy ${\cal E}=\langle\mathcal{F}\rangle+(\mu-\epsilon/2)n$ is then
\begin{equation}
{\cal E}=\frac{\lambda_a n^2}{2}(1-x^2)^2+\frac{\epsilon n}{2}
(x^2-1)+\sqrt{2n^3}g x(1-x^2)
\label{mye}
\end{equation}
We define the dimensionless parameters \(\alpha=\lambda_an^{1/2}/2g\sqrt{2}\) and \(\beta=\epsilon/2g\sqrt{2n}\leq0\). For $\beta<-1$, as long as \(\alpha\) is not too negative, there are two extrema as a function of $x$: the boundaries $x=\pm1$ are local minima ($M$) and a maximum ($A_1$) lies between $x=0$ and $x=1$. However, if \(\alpha\) is reduced until \((3+16\alpha^2-8\alpha\beta)^3=27(1-4\alpha\beta)^2\), we find two additional local extrema; a minimum at \(A_2\) and a maximum at \(A_0\). At $\beta=-1$, the $x=-1$ point bifurcates, and for $\beta>-1$ it is a local maximum and the local minimum ($A_2$) is found in the region $-1<x<0$. Illustrative plots are shown in figure~\ref{fig:energy}(a).

Previous analyses \cite{romans:qpt, radzihovsky:qpt} show that the MC
state $M$ is always stable against density fluctuation, and is
(thermodynamically and dynamically) stable against pairing
fluctuations if and only if $\beta<-1$.

Thermodynamic stability of the ASF is explored by calculating the Hessian $H_{ij}=\partial^2 {\cal E}/\partial i\partial j$, where $i,j=x,n$.  Using the condition $\partial {\cal E}/\partial x=0$, these derivatives can be written as $H_{xx}=n[\epsilon+2\lambda(4
n_m-n_a)-12 g\phi_m]$, $H_{xn}=H_{nx}=[g (4 n_m-n_a)-2\epsilon \phi_m]/\sqrt{2 n}$, $H_{nn}= n_a (2 n_a \lambda + 3 g
\phi_m)/(2n^2)$. The determinant of the Hessian (the discriminant) is related to the compressibility, \(\partial\mu/\partial n=(H_{nn}H_{xx}-H^2_{nx})/H_{xx}\). For $A_2$ the discriminant is always negative, while for $A_1$ and \(A_0\) it is negative for
$\lambda\epsilon>2g^2$ and otherwise positive.  Thus $A_2$, which is always stable against pairing fluctuations (\(H_{xx}>0\)), is always thermodynamically unstable towards density fluctuations (i.e. has a negative compressibility).  Similarly \(A_1\) and \(A_0\) are always thermodynamically unstable against pairing fluctuations (\(H_{xx}<0\)), and are thermodynamically unstable against density
fluctuations if and only if $\lambda\epsilon>2g^2$. 

Dynamical stability is explored by calculating the equations of motion for the fluctuations.  We write the field operators in terms of
density fluctuation $\hat\rho(r)$, pairing fluctuation $\hat y(r)$, relative phase fluctuation $\hat\chi(r)$, and total phase fluctuation $\hat \theta(r)$.
\begin{align}
\hat \psi_m(r) &= \sqrt{\frac{n+\hat\rho(r)}{2}}[x+\hat y(r)]e^{2i[\hat \theta(r)+\xi+\hat \chi(r)]}\\
\hat \psi_a(r) &= \sqrt{n+\hat\rho(r)}\sqrt{1-(x+\hat y(r))^2}e^{i[\hat
\theta(r)-\xi-\hat \chi(r)]}\notag
\end{align}
The equations of motion are found by making stationary the action
\begin{equation}
S=\int i \hat{\psi}_a^\dagger \partial_t\hat{\psi}_a+i \hat{\psi}_m^\dagger
\partial_t\hat{\psi}_m-{\cal F}
\end{equation}
Working to quadratic order in the fluctuations, we find
\begin{align}
&\dot{\rho}_k =
\frac{n}{m}k^2\theta_k-\frac{nu}{m}k^2\chi_k\label{eq:dynamicalmatrix}\\
&u\dot{\rho}_k-4nx\partial{y}_k =
\frac{nu}{m}k^2\theta_k-\left[\frac{n}{m}k^2+H_{\xi\xi}\right]\chi_k\notag\\
&nx\dot{\chi}_k =
\left[vk^2-H_{nx}\right]\frac{\rho_k}{4}-\left[\frac{n(3x^2+1)}{4m(1-x^2)}k^2+H_{xx}\right]\frac{y_k}{4}\notag\\
&\dot{\theta}_k-u\dot{\chi}_k =
\left[\frac{3x^2-4}{16mn}k^2-H_{nn}\right]\rho_k+\left[vk^2-H_{nx}\right]y_k\notag
\end{align}
where \(H_{\xi\xi}=-16\sqrt{2n^3}gx(1-x^2)\), \(u=1-2x^2\), \(v=3x/8m\), \(\dot{a}\equiv\partial_ta\) and the Fourier components
of the fluctuation operators are defined by ${O(r)=\sum_k O_k e^{ikr}}$.  As $k\to0$ the density and pairing modes decouple, and their frequencies are
\beq
\begin{split}
\omega_{\rm density}^2 &= c_s^2 k^2+{\cal O}(k^4)\\
\omega_{\rm pair}^2 &= \Delta^2 +{\cal O}(k^2)
\end{split}
\label{eq:gapandvelocity}
\eeq
where the speed of sound is related to the compressibility by the standard expression $c_s^2=(n/m)\partial\mu/\partial n$, and the gap to pairing excitation is \(\Delta^2=H_{xx}H_{\xi\xi}/16n^2x^2\). Since \(H_{\xi\xi}\propto -x\), \(A_1\) and \(A_2\) are dynamically stable against pairing fluctuations, while \(A_0\) is unstable. Conversely, we see a dynamic instability towards density fluctuations if and only if a thermodynamic instability exists.

\subsection{Effect of non-zero \(\lambda_m\) and \(\lambda_{am}\)}
We have seen that in the absence of \(\lambda_{am}\) and \(\lambda_m\) there is no stable ASF, and the metastable ASF always has larger energy than the MSF. Hence there is no MSF\(\leftrightarrow\)ASF phase transition. We now show the existence of a continuous
MSF\(\leftrightarrow\)ASF phase transition when \(\lambda_m>0\). To produce such a continuous phase transition it is necessary and
sufficient to show that there exists a stable ASF, with arbitrarily small atomic fraction, at the point the MSF becomes destabilized. In
the presence of a non-zero \(\lambda_m\) and \(\lambda_{am}\) figure~\ref{fig:energy}(b) represents the generic structure of
\(\mathcal{E}\); two minima at \(A_2\) and \(A_3\) and a maximum at \(A_1\). In terms of dimensionless parameters
\(\gamma=\lambda_m\sqrt{n/2}/8g\) and \(\eta=\lambda_{am}\sqrt{n/2}/2g\), \(A_{2,3}\) appears at \(x=\mp 1\) when \(\beta+2\gamma-\eta=\mp 1\), where the upper signs correspond to \(A_2\) and the lower signs correspond to \(A_3\).

The compressibility at \(x=\mp 1\) when \(A_{2,3}\) first appears is proportional to \(16\alpha\gamma-(1\mp 2\eta)^2\). So neither ASF is stable if \(\gamma=0\), i.e., even when \(\lambda_{am}\neq0\), a continuous MSF\(\leftrightarrow\)ASF phase transition cannot exist if \(\lambda_m=0\).

The curvature (\(H_{xx}\)) at \(x=\mp 1\) when \(A_{2,3}\) first appears is proportional to \(\pm 1+2(\alpha+\gamma-\eta)\). At
\(A_3\), \(H_{\xi\xi}\propto -x\) is negative, and therefore \(\Delta^2\propto H_{xx}H_{\xi\xi}\) is negative whenever \(H_{xx}>0\); i.e., \(A_3\) is always either dynamically or thermodynamically unstable against pairing fluctuations. The dynamical instability of \(A_3\) even when \(H_{xx}>0\) can be understood as instability against fluctuations in \(\xi\), i.e., in the \(x-\xi\) plane, the energy has a saddle-point at \(A_3\) (recall that \(4\xi\) is the relative phase between the atomic and molecular components). At \(A_2\), however, \(H_{xx}>0\) is equivalent to \(\Delta^2\propto H_{xx}H_{\xi\xi}>0\).

When the atom-molecule scattering vanishes (\(\eta=0\)), the stability conditions at \(A_2\), viz. \(H_{xx}>0\) and \(\partial\mu/\partial
n>0\) are simultaneously satisfied if and only if \(\gamma>0\) and \(16\alpha\gamma>1\). Thus there exists an MSF\(\leftrightarrow\)ASF continuous phase transition when \(\lambda_{am}=0\) iff \(n\lambda_a\lambda_m>2g^2\) and \(\lambda_m>0\) \footnote{The general criterion for the existence of a continuous MSF\(\leftrightarrow\)ASF phase transition when
\(\lambda_m\neq0\), \(\lambda_{am}\neq0\) can be worked out in the \(\lambda_m-\lambda_{am}\) space from the conditions
\(16\alpha\gamma>(1-2\eta)^2\) and \(1+2(\alpha+\gamma-\eta)>0\).}.
\newpage
\subsection{Stationary States (fixed chemical potential)}
Working at fixed chemical potential (and taking \(\lambda_{am}=0\)), there are two type of stationary points of Eq. (\ref{eq:freeenergy}); an MSF: \(\phi_a=0\), \(\phi_m^2=(2\mu-\epsilon)/\lambda_m\), and an ASF: \(\lambda_a\lambda_m\phi_m^3 + (\lambda_a(\epsilon-2\mu)-2g^2)\phi_m + \mu g=0\), \(\phi_a^2=(\mu-2g\phi_m)/\lambda_a\). The ASF equation has three solutions, one of which can be ruled out\cite{endnote21}. Stability analysis is done for both ASF and MSF states by considering fluctuations in \(\phi_m\), \(\phi_m^*\), \(\phi_a\) and \(\phi_a^*\), analogous to Eq. (\ref{eq:dynamicalmatrix}). In terms of dimensionless quantities \(\varphi=\phi_m\lambda_m/g\), \(r=\lambda_m/\lambda_a\), \(\varepsilon=\epsilon\lambda_a/g^2\) and \(\nu=\mu\lambda_a/g^2\), the MSF solution first appears at \(2\nu=\varepsilon\) and is stable for \(\nu<(4-2\sqrt{4-\varepsilon r})/r\). The two physical ASF solutions exist for \(\nu>\nu_c\) where \(4(\varepsilon-2\nu_c-2)^3+27\nu_c^2r=0\); one of them is always stable, the other is stable for \(\nu>0\). The ASF\(\leftrightarrow\)MSF tricritical point is obtained by demanding that the two physical ASF solutions appear exactly when the MSF destabilizes. Mathematically,
\begin{align}
4(\varepsilon-2\nu-2)^3+27\nu^2r &= 0\notag\\
\varphi^3+\varphi r(\varepsilon-2\nu-2)+\nu r^2 &= 0\\
\varphi^2-r(2\nu-\varepsilon) &=0\notag
\end{align}
Solving these three simultaneously gives the tricritical point \(\nu_{tc}=-2/\sqrt{r}\) and \(\varepsilon_{tc}=-1-4/\sqrt{r}\); \(r\) therefore uniquely determines the phase diagram. Coupled with \(n=2\phi_m^2=2(2\mu-\epsilon)/\lambda_m\), this yields the familiar result \(n\lambda_a\lambda_m=2g^2\).

\subsection{Discussion}
We have shown that a continuous ASF\(\leftrightarrow\)MSF phase transition can occur at sufficiently high density in a Bose gas near a
Feshbach resonance with repulsive molecule-molecule interaction. This ASF does not, however, correspond to the phase currently studied in cold atom experiments. The experimental ``phase'' is a saddle point of the free energy, and always has a higher energy than the MSF.

The most obvious route to studying this transition would involve first creating an MSF (for instance, using the technique of Xu {\it et
al} \cite{xu:msf}), then slowly ramping toward the resonance (making \(\mod{\epsilon}\) smaller). As pointed out by previous authors
\cite{romans:qpt, radzihovsky:qpt} the transition could be detected by observing the behavior of vortices.

We caution that this transition does not occur at arbitrarily low densities, nor in the absence of molecule-molecule scattering\footnote{Estimating \(\lambda_m\sim 4\pi\hbar^2 a_s/2m\) far from resonance, we see that in current experiments \cite{xu:msf, claussen:dynamics, donley:coherence} \(n\ll 2g^2/\lambda_a\lambda_m\), making observation of this phase transition impossible (\(g^2\sim(4\pi\hbar^2/m)a_{\text{bg}}\Delta\mu\Delta B\) \cite{mukaiyama:sodium}).}. Therefore inelastic 3-body processes will inevitably limit the lifetime of the cloud \cite{petrov:liquid}, perhaps making these experiments impractical. In fact, estimating the time scale of three-body recombinations\cite{braaten:threebody} to be \(\tau_{\text{3-body}}\sim m/\hbar a_s^4 n^2\) with \(a_s=\hbar/\sqrt{m\epsilon_c}\) (\(\epsilon_c\) being binding energy for the transition) and \(n=2g^2/\lambda_a\lambda_m\) already gives \(\tau_{\text{3-body}}\sim10^{-4}\text{s}\). Quantum interference effects can drastically reduce this decay rate, but only at particular binding energies\cite{braaten:threebody}. Using a photoassociation transition in lieu of Feshbach resonance may provide sufficient control over the parameters of the system to avoid these difficulties\cite{heinzen:superchemistry}.

This work was partially performed at the Aspen Center for Physics and was supported by the National Science Foundation (NSF) under grant PHY-0456261 and the Cornell Center for Material Research (DMR-0079992).

\end{document}